\documentclass[apj]{emulateapj}
\usepackage{natbib,apjfonts}

\shortauthors{KEMPNER \& DAVID}
\shorttitle{CHANDRA STUDY OF THE CORE OF A576}

\slugcomment{Accepted for publication in the Astrophysical Journal.}

\begin{document}

\title{A {\it Chandra} Study of the Core of the Nearby Cluster Abell 576}

\author{Joshua C. Kempner and Laurence P. David}
\affil{Harvard-Smithsonian Center for Astrophysics, 60 Garden St.,
Cambridge, MA 02138}
\email{jkempner@head-cfa.harvard.edu,
ldavid@head-cfa.harvard.edu}

\begin{abstract}
We present data from a {\it Chandra} observation of the nearby cluster of
galaxies Abell 576.  The core of the cluster shows a significant departure
from dynamical equilibrium.  We show that this core gas is most likely the
remnant of a merging subcluster, which has been stripped of much of its
gas, depositing a stream of gas behind it in the main cluster.  The
unstripped remnant of the subcluster is characterized by a different
temperature, density and metalicity than that of the surrounding main
cluster, suggesting its distinct origin.  Continual dissipation of the
kinetic energy of this minor merger may be sufficient to counteract most
cooling in the main cluster over the lifetime of the merger event.
\end{abstract}


\section{Introduction}
\label{sec:intro}

The study of the cores of clusters of galaxies has undergone a renaissance
in the past few years with the launch of the {\it Chandra} and {\it
XMM-Newton} observatories.  While previous observatories lacked the spatial
resolution necessary to resolve structure within the cores of clusters,
this new generation of telescopes has revealed an astonishing level of
complexity in the structure of the intracluster medium (ICM).  Many
clusters previously thought to be relaxed, regular systems have proven to
be far from dynamical equilibrium, particularly in their cores
\citep*[e.g.][]{mvm01,mem03}.  Abell 576 is no exception.  Earlier data,
especially optical spectra of the cluster's galaxy population, hinted at
dynamical complexity in the cluster core \citep{mgf+96}, but earlier X-ray
observations showed the cluster to be quite regular and even cooler in the
center, suggesting either a cooling flow or multi-phase gas with a very
cool component \citep{rvm+84,mgf+96}.

With its low redshift, Abell 576 makes excellent use of the capabilities of
{\it Chandra}, allowing us to examine in detail the very core of the
cluster.  In this paper, we focus on the dynamical activity in the core of
cluster.  The cluster shows strong evidence, first suggested by
\citet{mgf+96} from an analysis of the galaxy population, of the remnant
core of a small merged subcluster.  We demonstrate that the X-ray data are
consistent with this picture, and even suggest it as the most likely origin
for the non-equilibrium gas at the center of Abell 576.  In fact, the
subcluster may still be in the process of settling into the center of the
main cluster's potential.

Throughout this paper, we use the cosmological parameters derived from the
first release WMAP results \citep{bhh+03}, so $1\arcsec=0.738$ kpc at
$z=0.0377$.  All errors are quoted at 90\% confidence unless otherwise
stated.

\section{Observation and Data Reduction}
\label{sec:obs}

The data were obtained during {\it Chandra} Cycle 3 in a single exposure of
38.6 kiloseconds.  The focus was set on the back-illuminated S3 chip,
although significant flux from the cluster is detected on the adjacent
front-illuminated S2 chip.  The standard background reduction for Very
Faint (VF) mode data was applied using the CIAO tool acis\_process\_events.
In addition, the data were corrected for charge transfer inefficiency (CTI)
using the CXC/MIT CTI-corrector in CIAO version
2.3\footnote{http://asc.harvard.edu/ciao/}, although this only corrects the
front-illuminated chips such as the S2.  They were then filtered on the
standard {\it ASCA} grades 0, 2, 3, 4, and 6.  The data were processed
using version 2.18 of the {\it Chandra} CALDB.  CTI-corrected blank sky
background files\footnote{http://hea-www.harvard.edu/$\sim$maxim/axaf/acisbg/}
provided by Maxim Markevitch were used for the background correction.  The
VF mode background reduction was also applied to the background files.

The final 10.7 ksec of the observation was contaminated by a background
flare.  The flare was of the hard spectrum variety, which affects both the
front- and back-illuminated chips.  Unlike the soft flares that {\it
Chandra} sometimes experiences, which affect only the back-illuminated
chips, the spectra of these hard flares are not consistent from flare to
flare, and therefore cannot be modeled \citep{mar02}.  We therefore excised
the contaminated data, and only consider here the first 27.9 ksec of the
observation.

Because of the small field of view of the S3 chip and the low redshift of
Abell 576, our analysis is mostly restricted to the central $\sim$200 kpc
of the cluster.

For the spectroscopic analysis we considered only data in the range
0.5--8 keV.  Below this range the calibration of the ACIS CCDs is less
certain, and above this range the data contain few photons and are
dominated by particle background.  Spectral response matrices were
corrected for the reduction in quantum efficiency at low energies using
the time-dependent but spatially invariant {\it acisabs} model provided
by the Chandra X-ray Center.  Because of spatial variations in the
contaminant, the correction is not exact, especially below 1.2 keV
where the effect increases.  For this reason, we kept the absorption
column fixed to the Galactic value of $5.71\times10^{20}$~cm$^{-2}$
\citep{dl90} in the spectral fits presented here.  As a test, we
allowed the absorption column to vary, and measured a column density
nearly identical to the Galactic value, but with substantial errors,
typically on the order of 10\%.  The only effect of allowing the
absorption to be a variable parameter was to increase the errors in our
determinations of the other free parameters.  Thus, we concluded that
it was preferable to leave the absorption fixed at the Galactic value.

\begin{figure}
\epsscale{0.65}
\plotone{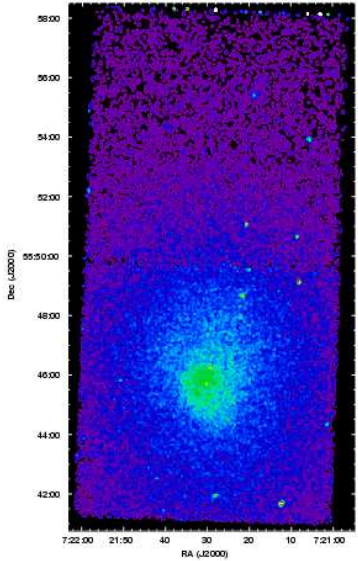}
\epsscale{1.0}
\caption{Gaussian smoothed, exposure-corrected, 0.3--6.0 keV image of
Abell 576.  Both the S2 and S3 chips are shown.  The north and southeast
surface brightness edges are clearly visible; the west edge is less
distinct.
\label{fig:img}}
\end{figure}

\section{Brightness Edges}
\label{sec:edges}

Figure~\ref{fig:img} shows a Gaussian smoothed, exposure-corrected
image of the cluster.  At least two, and perhaps more, surface
brightness edges are visible within the central 50 kpc.  As shown
below, they encompass a region of cool, high-metalicity gas.  This cool
gas also extends in a finger to the north of the cluster core, slightly
west of center.  As we will discuss in the remainder of this section,
we believe this finger of gas to have originated in a small subcluster
which is currently accreting into the center of the main cluster.  The
orientation of the edges are not consistent with gas simply sloshing
back and forth in a more or less fixed potential, which would create
parallel edges \citep*{qbb01} as opposed to the roughly triangular
configuration observed.  The observed edges are more consistent with
being the outer edges of a wake of stripped gas left behind by a
merging subcluster.  In this picture, the subcluster initially fell in
from the north, slightly to the west of the main cluster's center,
passed the main cluster center once, and is now making it's second pass
of the cluster's center.  The west and southeast edges describe the
outer edges of the wake of stripped gas from it's previous and current
pass of the main cluster's center.  This hypothesis also neatly
explains the finger of gas to the north, which cannot be easily
explained by simple sloshing.

\begin{figure}
\plotone{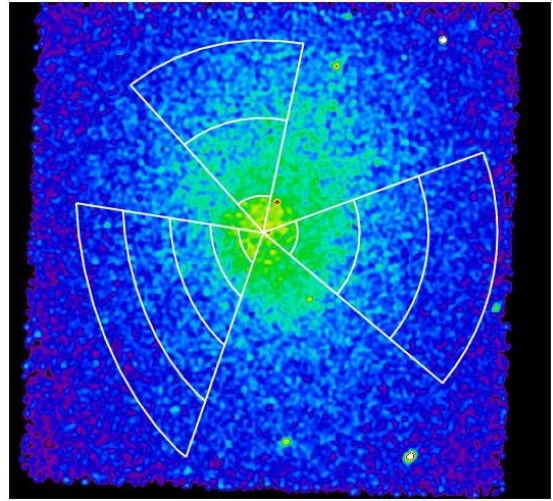}
\caption{Regions used for the spectral and surface brightness profiles.
The exact regions used for the spectral profiles are indicated.  The
surface brightness profiles used the same regions, but subdivided into
smaller radial bins.
\label{fig:regions}}
\end{figure}

\begin{figure}
\plotone{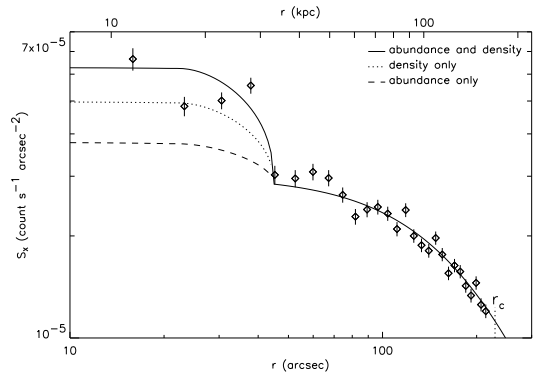}
\caption{Surface brightness profile across the north edge.  The solid line
outside 30 kpc is a beta model fit to the data using the parameters derived
by \protect\citet{mef+95} using the full field of view of the {\it
Einstein} IPC.  Inside 30 kpc, three models for the surface brightness jump
are shown: the solid line indicates the same beta model after adjusting for
the enhancement due to both the density and abundance increases inside the
edge.  The abundance only and density only components are shown as the
dashed and dotted lines, respectively .  For the $\beta$-model, $\beta =
0.64$ and $r_c = 169$ kpc.  The core radius of the model is indicated as
``${\rm r_c}$'' for reference.  The errors bars are 1$\sigma$.
\label{fig:north_sbr}}
\end{figure}

\begin{deluxetable*}{cccccccc}
\tablecaption{Spectral fit parameters outside the brightness edges
\label{tab:spec}}
\tablehead{
\colhead{} &
\colhead{$kT$} &
\colhead{($Z$)} &
\colhead{$\dot{M}$}&
\colhead{$kT_{\rm min}$} &
\colhead{$kT_2$} &
\colhead{} &
\colhead{} \\
\colhead{model} &
\colhead{(keV)} &
\colhead{($Z_\sun$)} &
\colhead{($M_\sun$ yr$^{-1}$)}&
\colhead{(keV)} &
\colhead{(keV)} &
\colhead{$\chi^2$} &
\colhead{d.o.f.}
}
\startdata
{\sc mekal}           & $4.05^{+0.17}_{-0.16}$ & $0.32^{+0.06}_{-0.07}$ & \nodata             & \nodata                & \nodata              & 357 & 297 \\
{\sc mekal + mkcflow} & $4.91^{+0.28}_{-0.51}$ & $0.37^{+0.08}_{-0.08}$ & $5.0^{+1.1}_{-1.0}$ & $0.01^{+0.20}_{-0.01}$ & \nodata              & 322 & 295 \\
{\sc mekal + mekal}   & $4.47^{+0.19}_{-0.19}$ & $0.47^{+0.09}_{-0.09}$ & \nodata             & \nodata                & $0.25^{+0.4}_{-0.2}$ & 289 & 295
\enddata
\end{deluxetable*}

\begin{figure}
\plotone{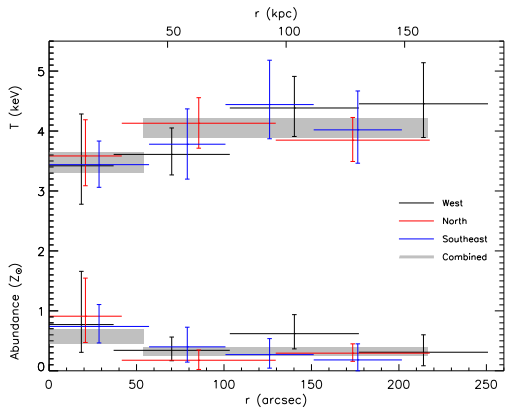}
\caption{Spectral profiles in three sectors to the north, southeast, and
west from the cluster center.  The top set of points are temperatures; the
bottom set are abundances.  All error bars are 90\% confidence.
\label{fig:spec_profile}}
\end{figure}

We extracted surface brightness and spectral profiles across all three
edges using the regions shown in Figure~\ref{fig:regions}.  The
brightness edge 40\arcsec\ north of the peak of the X-ray emission
shows by far the largest jump in surface brightness: a factor of
$1.8\pm0.15$ (1$\sigma$) increase across the discontinuity (see
Figure~\ref{fig:north_sbr}).  A large jump in the abundance is also
visible across the discontinuity, while the temperature does not change
significantly (see the points with red error bars in
Figure~\ref{fig:spec_profile}).  This jump in abundance is significant
at more than 90\% confidence across the north edge.  At the low
temperature of Abell 576, the increased abundance across the edge has a
non-negligible effect on the emissivity of the gas.  This is
illustrated in Figure~\ref{fig:north_sbr}, which shows the surface
brightness profile across the north edge.  Outside the brightness edge,
the solid line is a $\beta$-model fit to the data using the core radius
and slope determined from observations with {\it Einstein}
\citep{mgf+96}.  Inside the edge, we use the same $\beta$-model, but we
increase the emissivity by an amount expected from each of three
different models: the dotted line indicates the increased surface
brightness due to the increase in density alone; the dashed line
indicates the increased emissivity due to the higher abundance; and the
solid line is the increased brightness due to both effects.  For all
three models we assume spherical symmetry for consistency with the
deprojection analysis.  As the figure demonstrates, neither the added
emissivity from the higher abundance nor that from the increased
density can account entirely for the observed increase in surface
brightness, but the two effects combined reproduce the overall
normalization of the central brightness quite well.  We note that had
we kept the abundance in our spectral models fixed, we would have
underestimated the emissivity of the gas inside the brightness edge,
and would therefore have overestimated its density.  This would then
overestimate the density contrast across the edge, which would
negatively affect the analysis that follows.

We deprojected the surface brightness, which, when combined with the
temperature and abundance profiles, allowed us to determine deprojected
density and pressure profiles across the edge, under the assumption of
spherical symmetry.  We find a small jump in both the density and the
pressure across the north edge.  The density increases by a factor of
$2.8^{+0.8}_{-1.2}$ while the pressure increases by a factor of $2.4\pm0.8$
(both 1$\sigma$).  In order for the higher density gas to remain confined,
the pressure difference across the edge must be balanced by ram pressure
from motion of the high density gas through the lower density gas.  The
observed pressure difference implies that the higher density gas is moving
through the lower density gas with a velocity of $750 \pm 270$ km s$^{-1}$,
or Mach $0.9 \pm 0.3$ at the sound speed of the lower density gas (both
errors are 1$\sigma$).  We note that if we had failed to account for the
increase in abundance across the edge, that is, if we had assumed a
constant abundance on both sides of the edge, we would have overestimated
the density jump across the edge and consequently would have overestimated
the velocity of the dense gas cloud.  The velocity we measure is consistent
with velocities of both merging/accreting subclusters measured in other
clusters \citep[e.g.][]{mpn+00} and with velocities measured for some
``sloshing'' edges in the cores of relaxed cluster \citep{mar03}.  On its
own, then, the measured velocity of the north edge is incapable of
distinguishing between these two scenarios for the creation of the
non-hydrostatic features in the cluster core.

To the southeast of the cluster center, a fainter edge is visible in the
image (see Figure~\ref{fig:img}).  Another yet fainter edge appears to the
west.  Both of these edges display the same abundance gradient as the north
edge, though in both cases the abundance jump appears to be more of a
gradient than a sharp edge, and is measured with much less significance
(see Figure~\ref{fig:spec_profile}).

A combined spectral analysis of all 3 of sectors yields improved
statistics at the cost of spatial resolution.  This is plotted as the
solid grey boxes in Figure~\ref{fig:spec_profile}.  The best-fit models
for the combined spectra outside the edges are given in
Table~\ref{tab:spec}.  The columns are as follows: (1) spectral model;
(2) \& (3) temperature and abundance, respectively, of primary MEKAL
model; (4) Cooling rate of MKCFLOW model; (5) minimum temperature of
cool gas in MKCFLOW model; (6) temperature of second MEKAL model, where
the abundance of the second model is tied to that of the primary MEKAL
model; (7) \& (8) $\chi^2$ and number of degrees of freedom for the
fit.  From this combined analysis, we find that the central abundance
is different from the abundance outside the edges at greater than 90\%
confidence.  As expected, the temperature measured outside the edges is
approximately equal to the mean temperature measured with higher
spatial resolution in the individual sectors.  However, when fit with a
two temperature model, the best-fit high temperature is only slightly
higher than for the single temperature model, while the cool
temperature is $0.25^{+0.4}_{-0.2}$ keV and contributes only 5\% to
normalization of the model spectrum.  (Because of the errors on this
measurement and given that our spectral fit only includes data at
energies $>0.5$ keV, this cool component should be considered an upper
limit of $0.65$ keV.)  Similarly, a single temperature plus cooling
flow model yields a relatively small mass accretion rate of $5\pm1$
${\rm M_\sun~yr^{-1}}$ but to a very low temperature of less than $0.2$
keV.  An F-test shows that the two temperature model is a better fit
than the single temperature model at about the 3$\sigma$ level, while
the cooling flow model is better than the single temperature by only
$\sim$1$\sigma$.  In both multi-phase models, the cooler component
contributes a very small fraction of the total emission.

\begin{figure*}
\plottwo{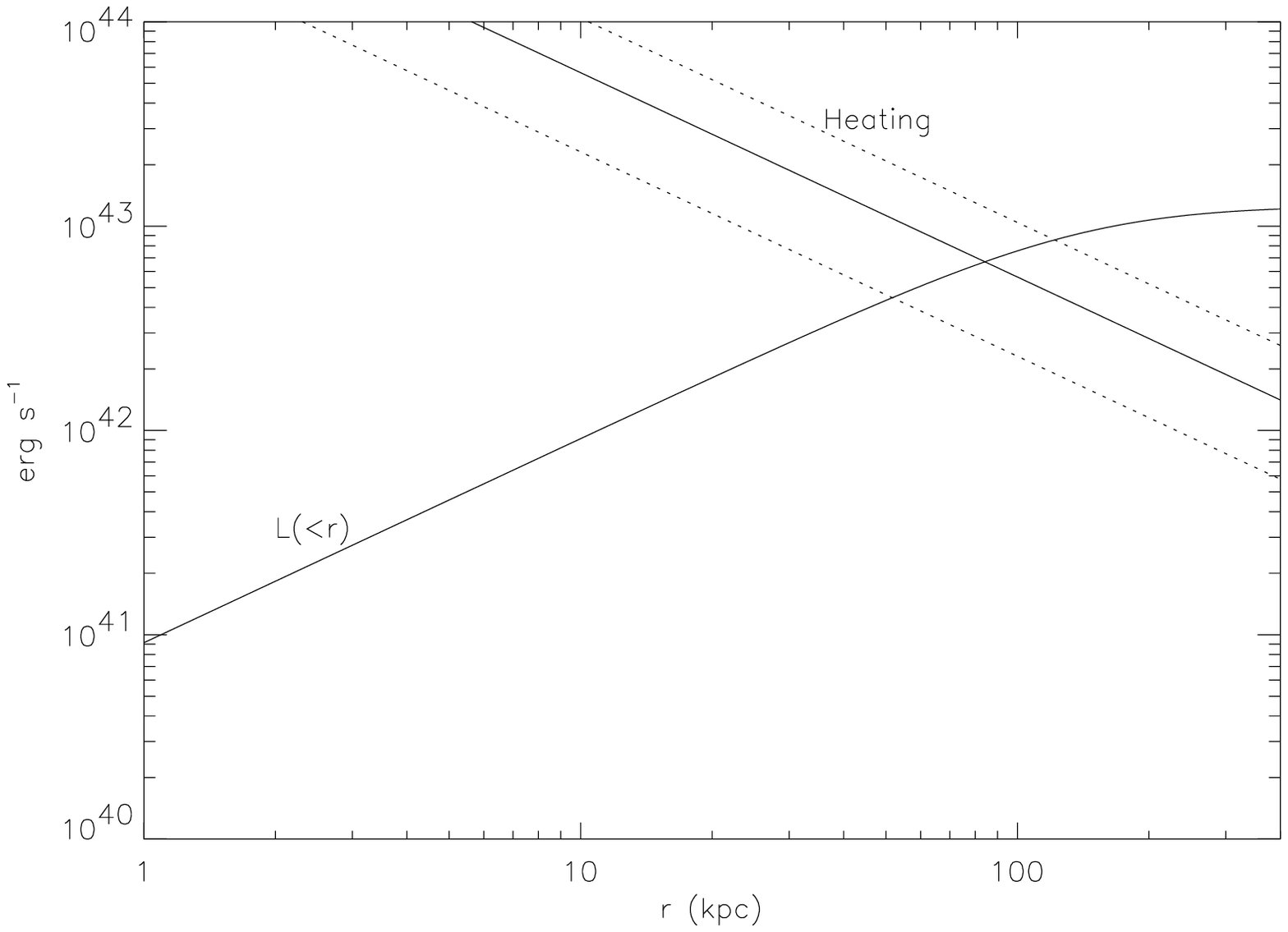}{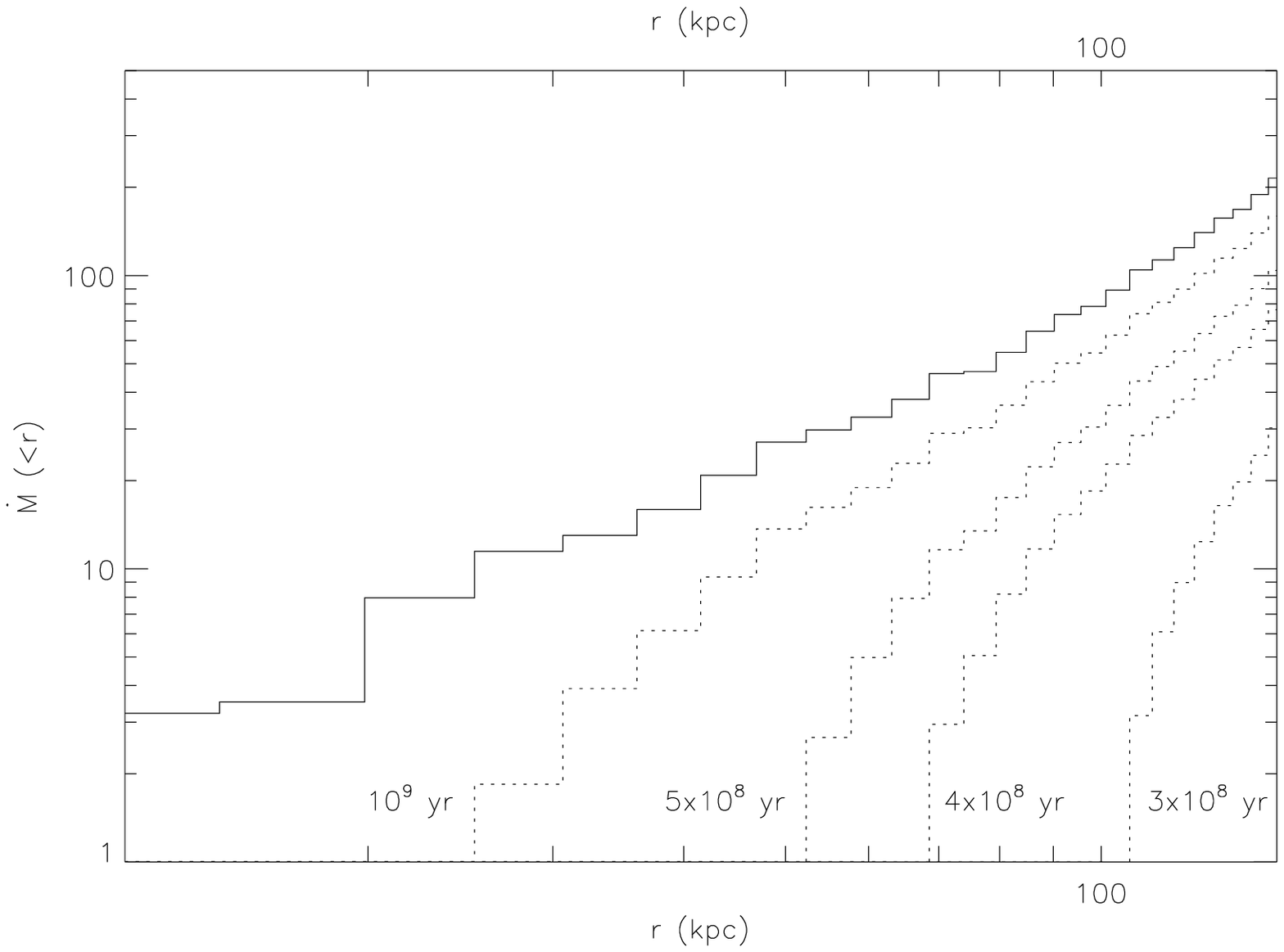}
\caption{{\it (a)} Heating rate and luminosity of radiative cooling as a
function of radius.  The exact definition of the heating rate is given in
the text.  The dotted curves indicate the minimum and maximum allowed rates
given the errors on our measurement of the subcluster's velocity.  {\it
(b)} Integrated cooling rate as a function of radius.  The solid line is
the rate in the absence of heating.  The dotted lines are the resulting
integrated cooling rate if the kinetic energy of the cool core is
dissipated over the given timescales.  The dissipated energy is assumed to
heat the gas at a constant rate per volume over the entire region.
\label{fig:cooling}}
\end{figure*}

\begin{deluxetable*}{cccccccc}
\tablecaption{Spectral fit parameters for cool core \label{tab:cflow}}
\tablehead{
\colhead{} &
\colhead{$kT$} &
\colhead{($Z$)} &
\colhead{$\dot{M}$}&
\colhead{$kT_{\rm min}$} &
\colhead{$kT_2$} &
\colhead{} &
\colhead{} \\
\colhead{model} &
\colhead{(keV)} &
\colhead{($Z_\sun$)} &
\colhead{($M_\sun$ yr$^{-1}$)}&
\colhead{(keV)} &
\colhead{(keV)} &
\colhead{$\chi^2$} &
\colhead{d.o.f.}
}
\startdata
{\sc mekal}           & $3.48^{+0.16}_{-0.18}$ & $0.56^{+0.13}_{-0.11}$ & \nodata     & \nodata               & \nodata                & 186 & 188 \\
{\sc mekal + mkcflow} & $3.77^{+0.25}_{-0.18}$ & $0.64^{+0.13}_{-0.11}$ & $1.1\pm0.5$ & $0.01^{+0.9}_{-0.01}$ & \nodata                & 183 & 186 \\
{\sc mekal + mekal}   & $3.62^{+0.25}_{-0.10}$ & $0.66^{+0.16}_{-0.14}$ & \nodata     & \nodata               & $0.65^{+2.11}_{-0.45}$ & 181 & 186 \\
{\sc mekal + foreground}& $3.21^{+0.27}_{-0.28}$ & $0.70^{+0.22}_{-0.19}$ & \nodata     & \nodata               & $4.05$\tablenotemark{a} & 186 & 188
\enddata
\tablenotetext{a}{\centering Temperature and abundance of foreground component taken from single-temperature fit in Table~\protect{\ref{tab:spec}}.}
\end{deluxetable*}

While a radial temperature gradient across the region used for our
analysis ($\sim$30--150 kpc) would show qualitatively similar effects,
we would expect such a gradient to cover a relatively small range of
temperatures, say, a factor of 2 as is seen in the most relaxed cooling
flow systems \citetext{e.g. \citealp{ppk+01}; \citealp*{bsm03}}.  To
find gas at a temperature less than $1/4$ of the ambient temperature is
quite unusual.  The merging subcluster hypothesis mentioned above
provides an attractive explanation for this extremely cool gas, if the
cool gas has been stripped from the subcluster during its infall.  The
small radius of curvature of the north edge suggests that this remnant
core is physically quite small and that the original subcluster was
also relatively small and therefore cool.  This  provides a consistent
qualitative explanation for the presence of the cool component.
Unfortunately, given the large amount of gas likely to have been
stripped from the subcluster during infall, it is not possible for us
to determine the subcluster's original mass and virial temperature and
is therefore not possible to conclude whether or not the cooler gas is
consistent with having originated in the undisturbed subcluster.
We note, however, that the ``sloshing'' picture for the brightness
edges cannot explain the existence of the cool component.  Sloshing
would create the same general appearance of a cold front
\citep[e.g.][]{mvm01}, but would not cause the higher density gas to
cool any more than it would in the absense of the sloshing.

An alternative explanation for the presence of gas at less than $1/4$
of the ambient temperature is that the usual method of preventing a
classical cooling flow from developing, whether that be AGN heating,
conduction, or some other mechanism, is absent or suppressed in A576.
Indeed the absence of the characteristic ``bubbles'' seen in other
clusters \citep[e.g.][]{fse+00,mwn+00,bsm+01} suggests there has been
no recent strong AGN activity, for one.  However, we find this
explanation to be unsatisfactory, since it would require the existence
of a strong cooling flow, which is explicitly ruled out by the small
cooling rate determined above.

\section{Cooling Flow}
\label{sec:cflow}

As Figure~\ref{fig:spec_profile} shows, the temperature drops in the very
core of the cluster, that is, inside the brightness edge.  It is natural to
ask, then, if this gas shows any evidence of being multi-phase.  To test
this, we fit a spectrum of the gas in the core with a single-temperature
absorbed MEKAL model \citep{kaa92}, with a MEKAL model plus a multi-phase
MEKAL model (MKCFLOW model), and with the sum of two MEKAL models.  The
salient parameters of the fits are shown in Table~\ref{tab:cflow}.  The
columns are as follows: (1) spectral model; (2) \& (3) temperature and
abundance, respectively, of primary MEKAL model; (4) Cooling rate of
MKCFLOW model; (5) minimum temperature of cool gas in MKCFLOW model; (6)
temperature of second MEKAL model, where the abundance of the second model
is tied to that of the primary MEKAL model; (7) \& (8) $\chi^2$ and number
of degrees of freedom for the fit.  As shown by the similar fit statistics,
the goodness of fit of all three models are essentially identical.

The spectroscopic cooling rate we measure is quite low: an order of
magnitude smaller than the ``classical cooling flow'' accretion rate of
$\dot{M} = M / t_{\rm cool} = 11$ M$_\sun$ yr$^{-1}$ derived from the gas
mass and cooling time in the inner 30 kpc (see Figure~\ref{fig:cooling} and
\S\ref{sec:suppress} below).  We note that the cool component in the
two-temperature model only contributes $\sim$0.1\% to the overall
normalization of the model.  This is consistent with the extremely small
cooling rate measured for the cooling flow model.

When the second temperature component of the two-MEKAL model was set to the
parameters of the fit to the outer gas (the equivalent of deprojection),
the fit was no better than for the single-temperature model.  This is
perhaps not too surprising, however, since the contribution of the
projected foreground component to the overall normalization of the model is
only about 2\%.

Surprisingly, the best-fit low temperature in the cooling flow model, that
is, the temperature to which the gas cools, is equal to the minimum
temperature allowed by the model, $kT_{\rm low} = 0.01$ keV.  However, the
error on the fit allows for $kT_{\rm low}$ to be as high as 0.9~keV.  This
is about 1/4 of the ambient temperature, which is consistent with minimum
temperatures found in stronger cooling flows \citep[e.g.][]{ppk+01}.  The
low temperature of this gas is consistent with that of the cool gas found
at larger radii.

In short, we find essentially no evidence of multi-phase gas in the
central cool core.  Of the models described above, the cooling flow
model contributes the largest contribution from multi-phase gas to the
overall emission, with only 5\% of the emission coming from the
multi-phase gas.

\section{Suppression of Cooling}
\label{sec:suppress}

Given that we observe essentially no cooling in either the core or outside
the core in what is otherwise a quite relaxed cluster, we now explore
whether the dissipation of the kinetic energy of the remnant core is
capable of suppressing cooling at the observed level.  In the rest of this
section we take the merging subcluster hypothesis to be the correct
explanation of the brightness edges.

We determined the kinetic energy of the gas inside the north edge using the
velocity of the edge we measured above plus a gas mass determined from the
deprojected density profile inside the edge ($\sim2\times10^{10}~{\rm
M_\sun}$).  We then calculated the rate of energy input from the
dissipation of this kinetic energy over a variety of timescales.
Figure~\ref{fig:cooling}a shows this energy dissipation rate compared to
the luminosity due to radiative cooling as a function of radius.  The
timescale used for calculating the heating rate is three crossing times of
the cluster to the given radius at the current velocity of the north edge.
In three crossing times, the moving cool core will have swept up its mass
in gas, reducing its kinetic energy by $3/4$.  We therefore assume
perfectly efficient thermalization of $3/4$ of the kinetic energy over a
timescale equal to three crossing times by the core at its current velocity
in calculating the heating rate.

If we take the point at which the west and southeast edges converge as
indicative of the current orbital radius of the subcluster ($\sim$100 kpc
from the cluster center), the dissipation timescale derived using the above
method is $4 \times 10^8$ years.  As can be seen from
Figure~\ref{fig:cooling}b, the heating simply from the dissipation of the
kinetic energy of the subcluster is capable of suppressing cooling by a
factor of more than 4 in the inner 100 kpc over this timescale.  If the
dissipation of the core's kinetic energy is spread out over $10^9$ years,
cooling can still be suppressed at the level observed in the inner 30 kpc.
Numerical simulations have shown that mergers of unequal mass clusters
thermalize their kinetic energy on timescales of one to a few times $10^9$
years \citep{rs01}, although the mass ratio involved in A576 is probably
much larger than has generally been tested in these simulations.

So far, we have omitted any discussion of the role of turbulence in
suppression of cooling.  Turbulence generated by the motion of such a small
subcluster would eventually dissipate and be thermalized on a timescale $<
10^8$ yr \citep*{fts03}, which is short compared to the timescale we are
considering for dissipation of the kinetic energy.  We therefore consider
it reasonable to assume close to 100\% thermalization efficiency.  We have
also neglected the influence of the dark matter on the total energy budget.
This is not so easily dismissed, as the dark matter halo of the subcluster
will take longer to dissipate its kinetic energy than will the gas.  The
dark matter oscillations about the center of mass of the system will induce
additional motion in the gas, transferring kinetic energy from the dark
matter to the gas and thereby increasing the total energy available for
heating the gas.  This will only serve to amplify the effect we have
discussed.  Our na\"ively calculated heating rate should therefore be taken
as a lower limit to the heating resulting from the subcluster merger.

Other alternatives for balancing cooling in clusters have been discussed.
These include heating from central AGN \citep{csf+02,bk02,fsa+03}, from
conduction \citep[e.g.][]{nm01}, and from turbulent mixing \citep{kn03}
induced by AGN activity.  Our data show no evidence for the X-ray cavities
typical of clusters with strong central AGN, and indeed the cluster has one
extremely weak AGN at its center with no non-nuclear radio emission.  Since
these cavities tend to persist for several AGN duty cycles
\citep[e.g. Perseus,][]{fse+00} and none are observed in A576, AGN heating
can be ruled out, at least over the last few times $10^8$ years.

\section{Conclusions}
\label{sec:conclusions}

While earlier X-ray observations of Abell 576 have shown it to be quite
regular on large scales, we have demonstrated that the core of the cluster
is far from dynamical equilibrium.  We found multiple surface brightness
edges in the cluster center which we have demonstrated to be indicative of
jumps in both density and abundance.  Of the two most likely explanations
for the existence of these edges, our analysis favors the hypothesis that
they are formed by gas stripped from a merging subcluster.  Most of the gas
appears to have been stripped from the subcluster, leaving a core only
$\sim$30 kpc in radius.  The stripped gas has been found to have both a
lower temperature and a higher abundance than the gas in the rest of Abell
576.

We find no evidence of gas cooling from the ambient temperature of the main
cluster, but do find some suggestion of very cool gas at a temperature
expected of gas that had condensed out of the ICM of the subcluster.  The
simple cooling rate derived from the gas mass and cooling time is an order
of magnitude larger than the spectroscopically measured $\dot{M}$.  We have
demonstrated that dissipation of the kinetic energy of the observed remnant
core of an infalling subgroup may be sufficient to reduce cooling to the
observed rate, if that energy is dissipated over a timescale of $\lesssim
10^9$ years.

\acknowledgements
Support for this work was provided by the National Aeronautics and Space
Administration through {\it Chandra} Award Number G01-2131X issued by the {\it
Chandra} X-ray Observatory Center, which is operated by the Smithsonian
Astrophysical Observatory for and on behalf of NASA under contract
NAS8-39073, and by NASA contract NAG5-12933.

\end{document}